\documentclass[12pt,english]{article}
\usepackage[latin9]{inputenc}
\usepackage{mathtools}
\usepackage{amsmath}
\usepackage{amssymb}

\makeatletter
\usepackage{babel}

\makeatother

\usepackage{babel}

\begin{document}
\title{$p$-Adic Polynomial Regression as Alternative to Neural Network for
Approximating $p$-Adic Functions of Many Variables}
\author{A.\,P.~Zubarev \\
 \textit{ Physics Department, Samara University, } \\
 \textit{ Moskovskoe shosse 34, 443123, Samara, Russia} \\
 \textit{Natural Science Department, } \\
 \textit{Samara State University of Railway Transport,} \\
 \textit{Perviy Bezimyaniy pereulok 18, 443066, Samara, Russia} \\
 e-mail:\:\texttt{apzubarev@mail.ru} }
\maketitle
\begin{abstract}
A method for approximating continuous functions $\mathbb{Z}_{p}^{n}\rightarrow\mathbb{Z}_{p}$
by a linear superposition of continuous functions $\mathbb{Z}_{p}\rightarrow\mathbb{Z}_{p}$
is presented and a polynomial regression model is constructed that
allows approximating such functions with any degree of accuracy. A
physical interpretation of such a model is given and possible methods
for its training are discussed. The proposed model can be considered
as a simple alternative to possible $p$-adic models based on neural
network architecture.

\textbf{Keywords:} $p$-adic alalysis, $p$-adic approximation, universal
approximating theorem, Mahler's interpolation series
\end{abstract}

\section{Introduction}

From the most general point of view, a neural network is a multiparameter
approximation of multidimensional functions of many real variables.The
structure of almost any neural network can be described by a directed
graph, each vertex of which has several incoming edges and one outgoing
edge. Directed edges of the graph correspond to real variables, and
vertices of the graph correspond to functions that map a set of variables
corresponding to incoming edges to a variable corresponding to an
outgoing edge. The functions corresponding to the vertices depend
on the parameters and the set of all these parameters forms the set
of parameters of the neural network. Thus, the neural network is a
model function $\boldsymbol{y}=\boldsymbol{f}_{nn}\left(\boldsymbol{x},\boldsymbol{w}\right)$,
which depends on the input parameters $\boldsymbol{x}$, takes values
on the space of output parameters $\boldsymbol{y}$ and is parameterized
by the set of parameters $\boldsymbol{w}$. The task of the neural
network is to approximate by the function $\boldsymbol{f}_{nn}\left(\boldsymbol{x},\boldsymbol{w}\right)$
an unknown (target) function $\boldsymbol{y}=\boldsymbol{f}\left(\boldsymbol{x}\right)$,
about which only some information is known, contained in the sample
arrays of input $\boldsymbol{X}$ and output $\boldsymbol{f}\left(\boldsymbol{X}\right)$
data. The solution to this problem consists of finding the values
of the parameters $\boldsymbol{w}$ (or, in the general case, a subdomain
of parameter values) for which the given function $\boldsymbol{f}_{nn}\left(\boldsymbol{x},\boldsymbol{w}\right)$
approximates the sample data $\left(\boldsymbol{X},\boldsymbol{f}\left(\boldsymbol{X}\right)\right)$
with a given degree of accuracy. To find the approximation $\boldsymbol{f}_{nn}\left(\boldsymbol{x},\boldsymbol{w}\right)$
it is necessary to solve the problem of minimizing the loss function
on the parameter space $\boldsymbol{w}$. The loss function can be
chosen in various ways, for example, in the form of the sum of squares
of norms of differences of the target function and values of the approximating
function on the training set of sample data $X_{train}$: $\underset{\boldsymbol{w}}{\min}\sum_{\boldsymbol{x}\in X_{train}}\left\Vert \boldsymbol{f}\left(\boldsymbol{x}\right)-\boldsymbol{f}_{nn}\left(\boldsymbol{x},\boldsymbol{w}\right)\right\Vert ^{2}$.
Generally speaking, since sample data are usually noisy, a random
component is usually added to the model function. Nevertheless, we
will not concern ourselves here with the statistical features of the
models and will consider the purely deterministic aspect of modeling.

Regression models have been known for a very long time, and like neural
networks, they are also designed to approximate multidimensional functions
of many real variables. From the most general point of view, regression
is a representation of a model function $\boldsymbol{y}=\boldsymbol{f}_{reg}\left(\boldsymbol{x},\boldsymbol{w}\right)$
in the form of some superposition of a set of given functions, parameterized
by a set of parameters $\boldsymbol{w}$. Just like for a neural network,
the task of regression is to approximate the function $\boldsymbol{y}=\boldsymbol{f}\left(\boldsymbol{x}\right)$
by a model function $\boldsymbol{f}_{reg}\left(\boldsymbol{x},\boldsymbol{w}\right)$
by finding the range of values of the parameters $\boldsymbol{w}$
for which this approximation is performed with a given accuracy.

Almost all regression and neural network models use real-valued multidimensional
arrays as input and output parameters and model parameters. Meanwhile,
a natural question arises: is it possible to use $p$-adic arrays
instead of real ones as input and output parameters and model parameters?
Such a need arises if the task is to approximate the target function
$\boldsymbol{y}=\boldsymbol{f}\left(\boldsymbol{x}\right)$, which
is not a real-valued function of several real variables, but a $p$-adic-valued
function of several $p$-adic variables. A similar problem can, in
principle, arise in a number of applications of $p$-adic analysis
and $p$-adic mathematical physics \cite{ALL,ALL_1}. It should be
noted that a number of works have been devoted to the application
of the $p$-adic approach to the construction of neural networks (see,
for example, \cite{Kh_1,Kh_2}). In these works, neural networks in
which the state of each layer is parameterized by a number from the
ring of $p$-adic integers were constructed and studied. Such models
are hierarchical in topology and can be considered as one of the effective
tools for solving pattern recognition problems.

As has been argued in a number of papers (see, for example, \cite{Kumar,Cheng,Morala})
almost any neural network model with any accuracy can be approximated
by a polynomial regression model. This follows from the fact that
any activation function in any compact domain can be approximated
with any degree of accuracy by a power series. On the other hand,
any polynomial regression model can also be approximated with any
degree of accuracy by a neural network model with activation functions
of sigmoid type, which follows from Cybenko's approximation theorem
\cite{Cyben,Liu}. Thus, polymial regression models and neural network
models are alternatives to each other. Nevertheless, as noted in \cite{Kumar,Cheng,Morala},
polynomial regression models can be much more effective than neural
network models in a number of problems involving approximation of
complex functions. In addition, the parameters of polynomial regression
models are better controlled than those of multilayer neural networks
when solving problems of multicollinearity and overtraining.

In this paper it is shown that, with respect to the problem of approximating
$p$-adic-valued functions of many $p$-adic variables, the polynomial
regression model arises naturally and is a simple alternative to other
possible $p$-adic models based on neural network architecture. The
paper is organized as follows. In Section 2, we present a method for
approximating functions $\mathbb{Z}_{p}^{n}\rightarrow\mathbb{Z}_{p}$
by a linear superposition of functions $\mathbb{Z}_{p}\rightarrow\mathbb{Z}_{p}$
($\mathbb{Z}_{p}$ is the ring of $p$-adic integers). In Section
3, we construct a polynomial regression model and give its possible
physical interpretation. In Section 4, we discuss possible methods
for training the proposed model.

\section{Approximation of function $\mathbb{Z}_{p}^{n}\rightarrow\mathbb{Z}_{p}$
by linear superposition of functions $\mathbb{Z}_{p}\rightarrow\mathbb{Z}_{p}$ }

Let us first present some information from $p$-adic analysis (see,
e.g., \cite{VVZ,Katok,KS}). Let $\mathbb{Q}$ be a field of rational
numbers and let $p$ be a fixed prime number. Any rational number
$x\neq0$ is representable as $x=p^{-\gamma}\dfrac{a}{b}$, where
$a$, $\gamma$ are integers, $b$ are natural number and $a$ and
$b$ are not divisible by $p$ and have no common multipliers. $p$-Adic
norm of the number $x\in\mathbb{Q}$ is defined as $\left|x\right|_{p}=p^{\gamma}$,
$\left|0\right|_{p}=0$. The completion of the field of rational numbers
by the $p$-adic norm forms the field of $p$-adic numbers $\mathbb{Q}_{p}$.
The metric $d\left(x,y\right)=\left|x-y\right|_{p}$ turns $\mathbb{Q}_{p}$
into a complete separable, totally disconnected, locally compact ultrametric
space. The ring of $p$-adic integers is $\mathbb{Z}_{p}=\left\{ x\in\mathbb{Q}_{p}:\:\left|x\right|_{p}\leq1\right\} $.
The canonical representation of a $p$-adic number $x\in\mathbb{Q}_{p}$
with the norm $\left|x\right|_{p}=p^{\gamma}$ is
\begin{equation}
x=p^{-\gamma}\sum_{i=0}^{\infty}x_{i}p^{i},\label{p_adic_ser}
\end{equation}
where $0\leq x_{i}\leq p-1$ for $i>0$ and $0<x_{0}\leq p-1$. The
$n$-dimensional space of $p$-adic numbers $\mathbb{Q}_{p}^{n}$
consists of points $\boldsymbol{x}=\left(x_{1},x_{2},\ldots,x_{n}\right)$,
where $x_{i}\in\mathbb{Q}_{p}$, $i=1,2,\ldots,n$. The $p$-adic
norm on $\mathbb{Q}_{p}^{n}$ is defined as $\left|\boldsymbol{x}\right|_{p}=\max_{i}\left(\left|x_{i}\right|_{p}\right)$.
This norm generates an ultrametric distance $d\left(\boldsymbol{x}^{(1)},\boldsymbol{x}^{(2)}\right)$
on $\mathbb{Q}_{p}^{n}$ for any two points $\boldsymbol{x}^{(1)},\boldsymbol{x}^{(2)}\in\mathbb{Q}_{p}^{n}$:
$d\left(\boldsymbol{x}^{(1)},\boldsymbol{x}^{(2)}\right)=\max_{i}\left(\left|x_{i}^{(1)}-x_{i}^{(2)}\right|_{p}\right)$.
The space $\mathbb{Q}_{p}^{n}$ (like $\mathbb{Q}_{p}$) is a complete
locally compact and totally disconnected ultrametric space. The space
$\mathbb{Z}_{p}^{n}$ is the set of points $\boldsymbol{x}=\left(x_{1},x_{2},\ldots,x_{n}\right)$,
where $x_{i}\in\mathbb{Z}_{p}$, $i=1,2,\ldots,n$. A $p$-adic-valued
function $f\left(\boldsymbol{x}\right)$ on $\mathbb{Q}_{p}^{n}$
( on $\mathbb{Z}_{p}^{n}$ ) is a map $\mathbb{Q}_{p}^{n}\rightarrow\mathbb{Q}_{p}$
($\mathbb{Z}_{p}^{n}\rightarrow\mathbb{Q}_{p}$). As in traditional
analysis, a function $f\left(\boldsymbol{x}\right)$ is said to be
continuous on $\mathbb{Q}_{p}^{n}$ (on $\mathbb{Z}_{p}^{n}$ ) if
for all $\boldsymbol{x},\boldsymbol{y}\in\mathbb{Q}_{p}^{n}$ (for
all $\boldsymbol{x},\boldsymbol{y}\in\mathbb{Z}_{p}^{n}$) we have
$\lim_{d\left(\boldsymbol{x},\boldsymbol{y}\right)\rightarrow0}\left|f\left(\boldsymbol{x}\right)-f\left(\boldsymbol{y}\right)\right|_{p}=0$.
On the set of continuous functions $f:\:\mathbb{Q}_{p}^{n}\rightarrow\mathbb{Q}_{p}(f:\:\mathbb{Z}_{p}^{n}\rightarrow\mathbb{Q}_{p})$
we can determine the norm $\left\Vert f\left(\boldsymbol{x}\right)\right\Vert _{p}=\underset{\boldsymbol{x}}{\max}\left|f\left(\boldsymbol{x}\right)\right|_{p}$.
A function $f:\:\mathbb{Q}_{p}^{n}\rightarrow\mathbb{Q}_{p}(f:\:\:\mathbb{Z}_{p}^{n}\rightarrow\mathbb{Q}_{p}{}_{p})$
is analytic if it is defined on $\mathbb{Q}_{p}^{n}$ (on $\mathbb{Z}_{p}^{n}$
) by a power series.

We will show that any continuous function $\mathbb{Z}_{p}^{n}\rightarrow\mathbb{Z}_{p}$
can be approximated with any degree of accuracy by a linear superposition
of continuous functions of one variable $\mathbb{Z}_{p}\rightarrow\mathbb{Z}_{p}$.
In this construction, the central place is occupied by the well-known
Mahler's theorem (see \cite{Mahler_1,Mahler_2,Bojanic}), which is
formulated here in the form of Theorem 1.

\textbf{Theorem 1.} Let $f:\:\mathbb{Z}_{p}\rightarrow\mathbb{Q}_{p}$
be a continuous function and let
\begin{equation}
w_{n}=\sum_{k=0}^{n}\left(-1\right)^{n-k}\binom{n}{k}f\left(k\right),\;n=0,1,2,\ldots.\label{a_n}
\end{equation}
Then the series
\[
\sum_{k=0}^{\infty}w_{k}\omega_{k}\left(x\right),
\]
where
\begin{equation}
\omega_{k}\left(x\right)=\binom{x}{k}\equiv\dfrac{x\left(x-1\right)\cdots\left(x-k+1\right)}{k!},\;x\in\mathbb{Z}_{p},\label{omega_k}
\end{equation}
converges uniformly on $\mathbb{Z}_{p}$ and
\begin{equation}
f\left(x\right)=\sum_{k=0}^{\infty}w_{k}\omega_{k}\left(x\right),\;x\in\mathbb{Z}_{p}.\label{f_decomp}
\end{equation}

Note that the decomposition (\ref{f_decomp}) also holds for any continuous
function $f\left(\boldsymbol{x}\right):\:\mathbb{Z}_{p}\rightarrow\mathbb{Z}_{p}$,
since $\left|\binom{n}{k}\right|_{p}\leq1$ for $n,k\in\mathbb{Z}_{+}$
and therefore $\left|a_{n}\right|_{p}\leq1$ for $f\left(x\right):\:\mathbb{Z}_{p}\rightarrow\mathbb{Z}_{p}$.
It is also easy to see that $\left|\omega_{n}\left(x\right)\right|_{p}\leq1$,
$x\in\mathbb{Z}_{p}$, $n\in\mathbb{Z}_{+}$ by $p$-adic continuity
of polinomials and its values on dence subset $\mathbb{Z}_{+}\subset\mathbb{Z}_{p}$
\cite{Mahler_2}.

From Mahler's theorem it follows that for any $\varepsilon>0$ there
exists $N\in\mathbb{Z}_{+}$ such that for all $n>N$ $\left|f\left(x\right)-\sum_{k=0}^{n}w_{k}\omega\left(x\right)\right|_{p}<\varepsilon$.
This means that any continuous function $f:\:\mathbb{Z}_{p}\rightarrow\mathbb{Z}_{p}$
can be approximated with any degree of accuracy by finite series of
the form $\sum_{k=0}^{n}w_{k}\omega_{k}\left(x\right)$ with $w_{k}\in\mathbb{Z}_{p}$.

Next, we will need the Theorem from \cite{Zubarev_2025}), which we
will formulate here in the form of Theorem 2.

\textbf{Theorem 2.} Let the function $f\left(x_{1},x_{2},\ldots,x_{n}\right):\:\mathbb{Z}_{p}^{n}\rightarrow\mathbb{Z}_{p}$
be continuous on $\mathbb{Z}_{p}^{n}$. Then it can be represented
as a superposition
\begin{equation}
f\left(x_{1},x_{2},\ldots,x_{n}\right)=h\left(\sum_{i=1}^{n}p^{i-1}\phi\left(x_{i}\right)\right),\label{main_2}
\end{equation}
where $h$ is a continuous function $h:\:\mathbb{Z}_{p}\rightarrow\mathbb{Z}_{p}$
and $\phi\left(x\right)$ is a continuous mapping $\mathbb{Z}_{p}\rightarrow\mathbb{Z}_{p}$
of the form

\begin{equation}
\phi\left(x\right)=\sum_{j=0}^{\infty}x_{j}p^{nj}.\label{omega}
\end{equation}

Since the function $h$ from Theorem 2 is continuous in $\mathbb{Z}_{p}$,
we can approximate it with any degree of accuracy by a finite decomposition
of the form (\ref{f_decomp}). In other words, for any continuous
function $h:\:\mathbb{Z}_{p}\rightarrow\mathbb{Z}_{p}$ and for any
$\varepsilon>0$, there exist such $K\in\mathbb{Z}_{+}$, $w_{k}\in\mathbb{Z}_{p}$,
$k=0,1,\ldots,N$ that inequality

\begin{equation}
\left\Vert h\left(x\right)-\sum_{k=0}^{K}w_{k}\omega_{k}\left(x\right)\right\Vert _{p}<\varepsilon\label{app_1}
\end{equation}
holds.

From (\ref{app_1}) and Theorem 2 follows a formula for approximating
any continuous function $f:\:\mathbb{Z}_{p}^{n}\rightarrow\mathbb{Z}_{p}$
by a linear superposition of continuous functions $\omega_{k}\left(\sum_{i=1}^{n}p^{i}\phi\left(x_{i}\right)\right)$
with any given accuracy, which we formulate in the form of Theorem
3.

\textbf{Theorem 3.} Let the function $f\left(x_{1},x_{2},\ldots,x_{n}\right):\:\mathbb{Z}_{p}^{n}\rightarrow\mathbb{Z}_{p}$
be continuous on $\mathbb{Z}_{p}^{n}$. Then for any $\varepsilon>0$
there exists such $K\in\mathbb{Z}_{+}$, $w_{k}\in\mathbb{Z}_{p}$,
$k=0,1,\ldots,K$ that the inequality
\begin{equation}
\left\Vert f\left(x_{1},x_{2},\ldots,x_{n}\right)-\sum_{k=0}^{K}w_{k}\omega_{k}\left(\sum_{i=1}^{n}p^{i-1}\phi\left(x_{i}\right)\right)\right\Vert _{p}<\varepsilon.\label{app_2}
\end{equation}
holds.

Note that despite the fact that the values of the functions $\omega_{k}\left(x\right)$
lie in $\mathbb{Z}_{p}$ and each of the functions $\omega_{k}\left(x\right)$
can be represented in the form of a polynomial of degree $k$ in the
variable $x$, the coefficients of these polynomials lie in $\mathbb{Q}_{p}$,
but not in $\mathbb{Z}_{p}$. For this reason, the approximation of
the function $f\left(x_{1},x_{2},\ldots,x_{n}\right)$ by a linear
superposition of power functions of the form $x_{i}^{k}$ with coefficients
from $\mathbb{Z}_{p}$ is not valid.

\section{Physical interpretation and model of $p$-adic polynomial regression}

The physical interpretation of the problem of approximating $p$-adic-valued
functions of several $p$-adic variables can be imagined as follows.
Let there be a set of $\mathcal{O}$ objects of the same type in the
number $N$ that we can observe and measure their characteristics.
Let us imagine that the observable characteristics of these objects
are more conveniently parameterized on the set of $p$-adic numbers
than on the set of real numbers. More specifically, we will assume
that to each object $X\in\mathcal{O}$ we can associate a set of measured
$p$-adic-valued characteristics, which we can divide into a set of
$n$ independent characteristics (input featires) $\boldsymbol{x}=\left(x_{1},x_{2},\ldots,x_{n}\right)\in M_{x}\subset\mathbb{Z}^{n}$
and the set of dependent characteristics (target variables) $\boldsymbol{y}=\left(y_{1},y_{2},\ldots,y_{m}\right)\in M_{y}\subset\mathbb{Z}^{m}$.
Thus, to each object with number $a$ we assign a pair $\left(\boldsymbol{x}^{(a)},\boldsymbol{y}^{(a)}\right)\in\mathbb{Z}^{n+m}$
and all these pairs form a dataset $D$. The main question we want
to pose is the following: are there quantitative relationships between
the characteristics of each object $\left(x_{1},x_{2},\ldots,x_{n}\right)$
and $\left(y_{1},y_{2},\ldots,y_{m}\right)$, expressed by a set of
$m$ ($m<n$) universal functional dependencies of the form $\boldsymbol{y}=\boldsymbol{f}\left(\boldsymbol{x}\right)$?

For simplicity, in what follows we will set $m=1$. We will assume
that the dependence expressed by the equation $y=f\left(\boldsymbol{x}\right):\mathbb{Z}_{p}^{n}\rightarrow\mathbb{Z}_{p}$
(the target function) exists but is unknown to us a priori. We also
assume that the function $f\left(\boldsymbol{x}\right)$ with a given
degree of accuracy can be pointwise approximated by a function from
the $K+1$-parameter function space $f\left(\boldsymbol{x},\boldsymbol{w}\right)$,
where $\boldsymbol{w}\in M_{w}\subset\mathbb{Z}_{p}^{K+1}$, $\boldsymbol{w}=\left\{ w_{i}\right\} $,
$i=0,1,2,\ldots,K$ and we can precisely define this space by explicitly
defining all functions $f\left(\boldsymbol{x},\boldsymbol{w}\right):\:\mathbb{Z}_{p}^{n+K+1}\rightarrow\mathbb{Q}_{p}$.
Traditionally, the function $f\left(\boldsymbol{x},\boldsymbol{w}\right)$
is called the model of the function $f\left(\boldsymbol{x}\right)$,
and the set $M_{w}$ is called the hypothesis space. Then the problem
of finding the function $f\left(\boldsymbol{x}\right)$ is reduced
to determining the parameters $\boldsymbol{w}=\boldsymbol{w}_{0}\in M_{w}$,
such that $f\left(\boldsymbol{x},\boldsymbol{w}_{0}\right)=f\left(\boldsymbol{x}\right)$
with a given degree of accuracy. This problem is equivalent to the
problem of minimization on the hypothesis space of the loss function,
which can be chosen in various forms. Here we choose a loss function
of the form

\begin{equation}
L\left(\boldsymbol{w}\right)=\dfrac{1}{N}\sum_{a=1}^{N}\left|l_{a}\left(\boldsymbol{w}\right)\right|_{p},\label{L_w}
\end{equation}
where
\begin{equation}
l_{a}\left(\boldsymbol{w}\right)=y^{(a)}-\sum_{k=0}^{K}w_{k}\omega_{k}\left(\sum_{i=1}^{n}p^{i-1}\phi\left(x_{i}^{(a)}\right)\right).\label{l_a_w}
\end{equation}
If the number of objects $N$ is rather large, then instead of the
space $\mathcal{O}$ of all objects, we can consider its subset $\mathcal{D}\subset\mathcal{O}$
with the number of elements $N^{D}<N$, which is a sample. Each sample
object $X_{a}\in\mathcal{D},a=1,\ldots,N^{D}$ has a set of measured
characteristics $\left(\boldsymbol{x}^{(a)},y^{(a)}\right)$. As in
the traditional machine learning approach, we divide the dataset $D$
into two subsets: $D^{train}\cup D^{val}=D$, where $D^{train}$ is
the training dataset (the sample of data used to fit the model) and
$D^{val}$ is the validation dataset (the sample of data used to provide
an unbiased evaluation of a model fit on the training dataset while
tuning model parameters $w$). Let the number of sets in $D^{train}$
and $D^{val}$ be $N^{train}$ and $N^{val}=N^{D}-N^{train}$, respectively.
Then instead of the function minimization problem (\ref{L_w}), we
can consider the problem of minimizing a function
\begin{equation}
L_{train}\left(\boldsymbol{w}\right)=\dfrac{1}{N^{train}}\sum_{a=1}^{N^{train}}\left|y^{(a)}-f\left(\boldsymbol{x}^{(a)},\boldsymbol{w}\right)\right|_{p},\label{L_D_train}
\end{equation}
on a training dataset $D^{train}$. As a model of the function $\boldsymbol{y}=\boldsymbol{f}\left(\boldsymbol{x}\right)$,
we choose the function
\begin{equation}
f\left(\boldsymbol{x},\boldsymbol{w}\right)=\sum_{k=0}^{K}w_{k}\omega_{k}\left(\sum_{i=1}^{n}p^{i-1}\phi\left(x_{i}^{(a)}\right)\right),\label{mod}
\end{equation}
which we will call the $p$-adic polynomial regression function and
which depends on $K$ parameters $w_{k}$. To approximate noisy data,
a random variable must also be added to the model function (\ref{mod})

\section{Discussion of the problem of training}

The problem of minimizing the loss function (\ref{L_w}) is fundamentally
different from the analogous problem in the real case. In the real
case, the loss function is usually a real-valued differentiable function
of many real variables and it is natural to apply gradient methods
to optimize it. For linear regression models, when the loss function
is represented in the form of the sum of squares of the norms of the
differences between the values of the target function and the values
of the approximating function, it is quadratic in the model parameters.
In this case, the values of the model parameters at which the loss
function reaches a minimum can be determined unambiguously. In the
$p$-adic case, the loss function (\ref{L_w}) is a real-valued function
of many $p$-adic variables, which is represented in the form of a
sum of $p$-adic norms of $p$-adic-valued functions. Such functions
are locally constant almost everywhere, so their derivatives on $p$-adic
variables are almost everywhere zero. Therefore, in the $p$-adic
case, alternative methods must be applied to optimize the loss function.

Obviously, the minimum possible value of the function (\ref{L_w})
is zero, which is achieved when the conditions
\begin{equation}
l_{a}\left(\boldsymbol{w}\right)=0\label{l_a_0}
\end{equation}
are met. In the case when $K=N-1$, $N=N^{train}$ the solution of
the system (\ref{l_a_0}) in $\mathbb{Q}_{p}^{N}$ is unique and trivial
to find. Indeed, in this case, since the functions $\omega_{k}\left(x\right)$
are linearly independent, the square matrix

\[
A_{ak}=\omega_{k}\left(\sum_{i=1}^{n}p^{i}\phi\left(x_{i}^{(a)}\right)\right),\;a=1,\ldots,N,\;k=0,1,\ldots,N-1
\]
has an inverse matrix $A_{ka}^{-1}$ and the solution of the system
(\ref{l_a_0}) are represented in the form

\begin{equation}
w_{k}=\sum_{a=1}^{N}A_{ka}^{-1}y^{(a)}.\label{sol_w_k}
\end{equation}
However, since solution (\ref{sol_w_k}) generally belongs to $\mathbb{Q}_{p}^{N}$,
a solution of equations (\ref{l_a_0}) in $\mathbb{Z}_{p}^{N}$ may
not exist. A necessary and sufficient condition for the existence
of a unique solution (\ref{sol_w_k}) in $\mathbb{Z}_{p}^{N}$ is
the existence of a point $\boldsymbol{w}^{(0)}=\left(w_{0}^{(0)},w_{1}^{(0)},\ldots,w_{N-1}^{(0)}\right)\in\mathbb{Z}_{p}^{N}$,
such that the inequality
\begin{equation}
\left|\boldsymbol{l}\left(\boldsymbol{w}^{(0)}\right)\right|_{p}\leq\left|A\right|_{p}^{2},\;A=\det\left(A_{ak}\right)\label{L<A}
\end{equation}
holds. Indeed, let's write it down
\[
w_{k}=\sum_{a=1}^{N}A_{ka}^{-1}y^{(a)}=w_{k}^{(0)}+\sum_{a=1}^{N}A_{ka}^{-1}l_{a}\left(\boldsymbol{w}^{(0)}\right).
\]
From (\ref{L<A}) it follows that

\[
\left|\sum_{a=1}^{N}A_{ka}^{-1}l_{a}\left(\boldsymbol{w}^{(0)}\right)\right|_{p}\leq\left|A\right|_{p}.
\]
Then from the last two relations we have

\[
\left|\boldsymbol{w}\right|_{p}\leq1.
\]

In the case where $K>N-1$, $N=N^{train}$ the system (\ref{l_a_0})
can have many solutions. Let us assume that for the fixed point $\left(w_{N}^{(0)},w_{N+1}^{(0)},\ldots,w_{K}^{(0)}\right)\in\mathbb{Z}_{p}^{K-N+1}$
there exists a point $\left(w_{0}^{(0)},w_{1}^{(0)},\ldots,w_{N-1}^{(0)}\right)\in\mathbb{Z}_{p}^{N}$,
such that the inequality $\left|\boldsymbol{l}\left(\boldsymbol{w}^{(0)}\right)\right|_{p}\leq\left|A\right|_{p}^{2}$,
$A=\det\left(A_{ak}\right)$, $k=0,\ldots N-1$ holds. In this case,
the first $N$ parameters $w_{k}$, $k=0,\ldots N-1$ can be expressed
through the remaining $K-N$ free parameters $w_{j},j=N,\ldots,K$.
However, there is no guarantee that the solution found with some fixed
$w_{j},j=N,\ldots,K$ will optimize the function (\ref{L_w}) on the
validation dataset. Therefore, it is very likely that this solution
will correspond to an overtrained model and additional optimization
on the solution space of the system (\ref{l_a_0}) is required. One
way of additional optimization can be to find a solution such that
the function $\sum_{k=0}^{N}\left|w_{k}\right|_{p}$ reaches a minimum
on the parameter space $w_{k}$, $k=N,\ldots,K$. This problem in
turn reduces to the optimization of the sum of norms of linear $p$-adic
functions with the number of parameters less than the number of summands
in this sum.

The case $K<N-1$ is the most interesting from a physical point of
view, because usually in regression and neural network models the
number of data in training sets is much larger than the number of
model parameters. In this case, the equations (\ref{l_a_0}), in general,
have no solution because the minimum possible value of the loss function
(\ref{L_w}) can be a positive rational number. Unfortunately, we
are not aware of a deterministic recurrence algorithm for finding
the minimum of such functions in the $p$-adic case.

From the discussion above it follows that an adequate deterministic
algorithm for training the model remains questionable. Nevertheless,
we can propose a stochastic optimization algorithm for the function
(\ref{L_w}) based on a random walk in the parameter space $\boldsymbol{w}$,
which is described by a non-Markovian random process at discrete time
$t=i=0,1,2,\ldots$:

\begin{equation}
\boldsymbol{w}_{i+1}=\boldsymbol{w}_{i}+\boldsymbol{\xi}_{i}\left(\boldsymbol{w}_{i},\beta_{i}\right).\label{proc}
\end{equation}
In the formula (\ref{proc}) $\beta_{i}$ is a sequence of positive
real numbers, $\boldsymbol{\xi}_{i}\left(\boldsymbol{w},\beta\right)$
is a sequence of independent random functions with values in $\mathbb{Z}_{p}^{K+1}$
depending on the random variable $\boldsymbol{w}\in\mathbb{Z}_{p}^{K+1}$
and also on the positive number $\beta$. Each of the functions $\boldsymbol{\xi}_{i}\left(\boldsymbol{w},\beta\right)$
has a conditional distribution law of the following form

\begin{equation}
d\mathrm{P}\left(\boldsymbol{\xi}\mid\boldsymbol{w},\beta\right)=\mathcal{N}^{-1}\left(\boldsymbol{w},\beta\right)\exp\left(-\beta\left(L\left(\boldsymbol{w}+\boldsymbol{\xi}\right)-L\left(\boldsymbol{w}\right)\right)\right)d_{p}^{K+1}\boldsymbol{\xi},\label{measure}
\end{equation}
where
\[
\mathcal{N}\left(\boldsymbol{w},\beta\right)=\intop_{\mathbb{Z}_{p}^{K+1}}d_{p}^{K+1}\boldsymbol{\xi}^{\prime}\exp\left(-\beta\left(L\left(\boldsymbol{w}+\boldsymbol{\xi}^{\prime}\right)-L\left(\boldsymbol{w}\right)\right)\right),
\]
and $d_{p}^{K+1}\boldsymbol{\xi}$ is the $K+1$-dimensional $p$-adic
Haar measure \cite{VVZ,KS}.

Next, we will show that under certain conditions the process (\ref{proc})
ensures a strict decrease of the loss function $L\left(\boldsymbol{w}_{i}\right)$
on average. Let $\boldsymbol{w}_{i}$ be a fixed realization of the
process (\ref{proc}) at the $i$-th step, parameterized by the continuity
of numbers $\beta_{j}$, $j=0,1,\ldots,i-1$ and we assume that the
set of points $U_{i}^{-}=\left\{ \boldsymbol{\xi}:\:L\left(\boldsymbol{w}_{i}+\boldsymbol{\xi}\right)-L\left(\boldsymbol{w}_{i}\right)<0\right\} $
has a non-zero measure $\intop_{U_{i}^{-}}d_{p}^{K+1}\boldsymbol{\xi}$.
Let us write down
\[
\intop_{\mathbb{Z}_{p}^{K+1}}L\left(\boldsymbol{w}_{i}+\boldsymbol{\xi}\right)d\mathrm{P}\left(\boldsymbol{\xi}\mid\boldsymbol{w}_{i},\beta_{i}\right)=
\]
\[
=L\left(\boldsymbol{w}_{i}\right)+\mathcal{N}^{-1}\left(\boldsymbol{w}_{i},\beta_{i}\right)\intop_{\mathbb{Z}_{p}^{K+1}}\exp\left(-\beta_{i}\left(L\left(\boldsymbol{w}_{i}+\boldsymbol{\xi}\right)-L\left(\boldsymbol{w}_{i}\right)\right)\right)\left(L\left(\boldsymbol{w}_{i}+\boldsymbol{\xi}\right)-L\left(\boldsymbol{w}_{i}\right)\right)d_{p}^{K+1}\boldsymbol{\xi}
\]
\begin{equation}
=L\left(\boldsymbol{w}_{i}\right)-\mathcal{N}^{-1}\left(\boldsymbol{w}_{i},\beta_{i}\right)\dfrac{\partial\mathcal{N}\left(\boldsymbol{w}_{i},\beta_{i}\right)}{\partial\beta_{i}}\label{PL>L}
\end{equation}
Note that $\mathcal{N}\left(\boldsymbol{w}_{i},0\right)=1$ and from
the assumption $\intop_{U_{i}^{-}}d_{p}^{K+1}\boldsymbol{\xi}>0$
it follows that

\[
\lim_{\beta_{i}\rightarrow\infty}\mathcal{N}\left(\boldsymbol{w}_{i},\beta_{i}\right)=\lim_{\beta_{i}\rightarrow\infty}\intop_{U_{i}^{-}}\exp\left(-\beta_{i}\left(L\left(\boldsymbol{w}_{i}+\boldsymbol{\xi}\right)-L\left(\boldsymbol{w}_{i}\right)\right)\right)d_{p}^{K+1}\boldsymbol{\xi}=\infty.
\]
Therefore, there is a range of values $\beta_{i}$ in which the function
$\mathcal{N}\left(\boldsymbol{w}_{i},\beta_{i}\right)$ increases
and $\dfrac{\partial\mathcal{N}\left(\boldsymbol{w}_{i},\beta_{i}\right)}{\partial\beta_{i}}>0$.
Then, taking into account (\ref{PL>L}), it follows that

\begin{equation}
\intop_{\mathbb{Z}_{p}^{K+1}}L\left(\boldsymbol{w}_{i}+\boldsymbol{\xi}\right)d\mathrm{P}\left(\boldsymbol{\xi}\mid\boldsymbol{w}_{i},\beta_{i}\right)<L\left(\boldsymbol{w}_{i}\right).\label{neq}
\end{equation}
According to inequality (\ref{neq}) it is possible to construct a
$K+1$-dimensional $p$-adic-valued random process of the form (\ref{proc})),
such that the random process $L\left(\boldsymbol{w}_{i}\right)$ is
a supermartingale in the strict sense, i.e. it ensures a strict decrease
of the function $L\left(\boldsymbol{w}_{i}\right)$ on average. Therefore,
based on the process (\ref{proc}), it is possible to generate a statistical
machine algorithm for minimizing the loss function (\ref{L_w}) in
the regression model (\ref{mod}).

\section*{Acknowledgments}

The study was supported by the Ministry of Higher Education and Science
of Russia by the State assignment to educational and research institutions
under project no. FSSS-2023-0009.

\section*{Data Availability Statement}

The data supporting the findings of this study are available within
the article and its supplementary material. All other relevant source
data are available from the corresponding author upon reasonable request.

\end{document}